\documentclass[pre,print,twocolumn,superscriptaddress,showpacs,aps,floatfix]{revtex4-1}
\setcitestyle{square}
\usepackage[utf8]{inputenc}
\usepackage{graphicx,url, microtype, graphicx, dcolumn, bm, dsfont, csquotes, dsfont, amsmath, amsbsy,amsfonts,appendix, booktabs, amssymb, mathtools,mathrsfs}
\usepackage[dvipsnames]{xcolor}
\usepackage[colorlinks,linkcolor=blue,citecolor=blue]{hyperref}
\usepackage{algorithm, algpseudocode}
\usepackage[utf8]{inputenc}
\usepackage[T1]{fontenc}
\usepackage[margin=1in]{geometry}
\usepackage[sort&compress]{natbib}
\usepackage[skins,theorems]{tcolorbox}
\usepackage{cleveref}
\usetikzlibrary{intersections, angles, fadings, through, positioning,arrows,shapes,automata,petri,positioning,calc}

\newcommand{\bs}[1]{{\boldsymbol{#1}}}

\mathchardef\mhyphen="2D
\bibstyle{aps}

\begin{document}
\title{Exo-Daisy World: Revisiting Gaia Theory through an Informational Architecture Perspective}
\author{Damian R Sowinski}
\email{Damian.Sowinski@Rochester.EDU}
\affiliation{Department of Physics and Astronomy, University of Rochester, Rochester, NY, 14627, USA}
\author{Gourab Ghoshal}
\email{gghoshal@pas.rochester.edu}
\affiliation{Department of Physics and Astronomy, University of Rochester, Rochester, NY, 14627, USA}
\affiliation{Department of Computer Science, University of Rochester, Rochester, NY, 14627, USA}

\author{Adam Frank}
\email{afrank@pas.rochester.edu}
\affiliation{Department of Physics and Astronomy, University of Rochester, Rochester, NY, 14627, USA}

\begin{abstract}
The Daisy World model has long served as a foundational framework for understanding the self-regulation of planetary biospheres, providing insights into the feedback mechanisms that may govern inhabited exoplanets. 
In this study, we extend the classic Daisy World model through the lens of Semantic Information Theory (SIT), aiming to characterize the information flow between the biosphere and planetary environment---what we term the \emph{information architecture} of Daisy World systems. 
Our objective is to develop novel methodologies for analyzing the evolution of coupled planetary systems, including biospheres and geospheres, with implications for astrobiological observations and the identification of agnostic biosignatures.
To operationalize SIT in this context, we introduce a version of the Daisy World model tailored to reflect potential conditions on M-dwarf exoplanets, formulating a system of stochastic differential equations that describe the co-evolution of the daisies and their planetary environment. 
Analysis of this Exo-Daisy World model reveals how correlations between the biosphere and environment intensify with rising stellar luminosity, and how these correlations correspond to distinct phases of information exchange between the coupled systems. 
This \emph{rein control} provides a quantitative description of the informational feedback between the biosphere and its host planet.
Finally, we discuss the broader implications of our approach for developing detailed ExoGaia models of inhabited exoplanetary systems, proposing new avenues for interpreting astrobiological data and exploring biosignature candidates.
\end{abstract}

\maketitle
\tableofcontents
\section{Introduction}\label{sec: introduction}

Understanding the long-term evolution of planetary biospheres is a crucial frontier in astrobiology. 
The search for life on exoplanets requires the identification of biosignatures, which rely on life having significantly altered the spectroscopic properties of a planet. 
Thus, exoplanetary life searches focus not on detecting individual organisms but on identifying the collective effects of life on the planetary system---what we refer to as exo-biospheres.

The study of biosignatures is therefore inseparable from the study of biospheres. 
Given that exoplanets may be observed at any point in their evolutionary history, it becomes essential to understand how and when biospheres reach a \emph{mature} state, wherein they exert strong feedback on the planetary geospheres (atmosphere, hydrosphere, cryosphere, and lithosphere). 
This same argument applies to \emph{technosignatures} and \emph{technospheres} (i.e., intelligence \cite{frank2022intelligence, zalasiewicz2017scale}) but in this work, we focus exclusively on non-technological life.

Vladimir Vernadsky provided the first comprehensive account of the term \emph{biosphere} \cite{vernadsky1945biosphere} by recognizing that life is not a passive presence but an evolutionary force in its own right. 
However, it was James Lovelock and Lynn Margulis who fully articulated the range of potential biospheric effects on a planet through their Gaia Theory \cite{lovelock1974atmospheric,margulis1974biological,lovelock1982regulation,onori2012gaia}. 
Initially termed \emph{Self-regulating Earth System Theory}, the Gaian model posited by Lovelock and Margulis emphasized the dense feedback loops between the biosphere and the abiotic planetary system, allowing a planet to maintain habitability over long timescales.

Gaia Theory faced considerable resistance upon its introduction, particularly regarding issues of teleology (Dawkins 1983 \cite{Dawkins1982-DAWTEP-2}) and natural selection (Doolittle 2017 \cite{DOOLITTLE201711}). 
While questions remain regarding full planetary homeostasis \cite{kirchner2002gaia}, the fundamental principles of Gaia Theory, repackaged as Earth Systems Science (ESS), now form the foundation of modern approaches to Earth's evolutionary history (Steffan et al. 2020 \cite{2020NRvEE...1...54S}). 
ESS adopted Gaia Theory's recognition of the biosphere as a key driver of planetary evolution, a concept that astrobiology also applies in the search for exoplanetary biosignatures. 
Progress in astrobiology thus depends on advancing our understanding of how biospheres co-evolve with planetary geospheres (atmosphere, hydrosphere, cryosphere, lithosphere). 
This requires untangling the complex network of Gaian (or semi-Gaian) feedbacks that regulate planetary systems. 
As a complex adaptive system (Krakauer 2024 \cite{Krakauer_complex}, the biosphere cannot be fully understood through reductionist approaches; its dynamics span multiple scales, with both upward and downward causal chains. 

In this work, we explore an alternative approach to characterizing biospheric complexity by combining physical descriptions of bio-geosphere feedback with information-theoretic measures, constructing what we call an \emph{information narrative}. 
To do so, we investigate the simplest model of biospheric homeostasis: the well-known Daisy World model \cite{watson1983biological}.
Daisy World is a toy model originally developed to illustrate Gaian self-regulation. 
The model consists of a planet with white and black daisies and a star with increasing luminosity over time. 
As the star's brightness changes, the daisy populations either grow or diminish, modifying the planet's albedo in such a way that its temperature remains within a habitable range for the daisies. 

We introduce a stochastic variant of Watson and Lovelock's model, which we refer to as Exo-Daisy World, to analyze aspects of its informational architecture. 
The original Daisy World model has been extensively studied and generalized in various ways, as highlighted in Wood's comprehensive 2008 review \cite{wood2008daisyworld} and more recently in Savi and Viola's 2023 variant \cite{savi2023daisyworld}. 
To our knowledge, our variant is the first to generalize Daisy World using coupled stochastic differential equations, which allow the application of information-theoretic measures to the model. 
Our overarching aim is to bring these information-centric methods to the study of biospheres, clarifying the processes underlying planetary co-evolution.

Our manuscript is organized as follows: in Section~\ref{sec: introduction}, we provide a brief overview of Semantic Information Theory, the information-theoretic framework we apply to the Daisy World model. 
In Section~\ref{sec: model}, we introduce our modified Daisy World model, incorporating stochastic variations in stellar luminosity. 
Section~\ref{sec: info_n} presents the information measures drawn from Semantic information theory, which we apply to exo-Daisy World. 
Section~\ref{sec: results} outlines our results and their implications, followed by our conclusions in Section~\ref{sec: discuss}. 
Technical aspects of our model, details of the numerics and the approximations used are outlined in Appendix~\ref{app: simulation}. 

\section{Semantic Information Theory and Informational Architecture}\label{sec: introduction} 

Over the past three decades, innovative research avenues to explore life's self-production and self-maintenance have emerged from fields such as far-from-equilibrium thermodynamics and network theory.  
Both these perspectives have helped capture the essential nature of life as an emergent complex system. More recently, a critical new dimension has been added to these studies by focusing on life as a physical system that uses information (i.e. storage, copying, transfer and computation)~\cite{ walker2016informational, egbert2018methods}.  
By marrying an information theoretic description with both network and thermodynamic constraints, life's information architecture emerges as a key conceptual construct for understanding self-production and self-maintenance i.e. agency and autonomy.   

Information's role in developing a {\it physics of life} faces a key challenge. Information theory, from Shannon's foundational work~\cite{shannon1948mathematical}, focuses on syntactic measures, defining information via the combinatorics of strings without addressing the {\it semantic content} of information. 
While useful in engineering, this approach omits how life utilizes information meaningfully~\cite{Schlosser_1998,Mossio_2009}. 
For instance, understanding agency revolves around a system's ability to process information for its survival, often linked to its {\it viability}, a critical factor distinguishing living systems~\cite{barandiaran2014norm,egbert2023behaviour}.

Although there are philosophical and computational discussions of semantic information~\cite{Polani_2001,Thompson_2009,nehaniv_meaningful_2003,barham1996dynamical,deacon2007shannon,corning2007control, gleiser2018we,gleiser2018we,sowinski2016complexity}, a mathematically rigorous theory applicable to diverse scientific fields, including the Origin of Life, has been missing~\cite{Acebron_2005, Garcia_2021, Cooper_2013, Mimar_2019, Cohen_1977, Rooney_2006,Xavier_2020, Craig_2014,Mimar:2021si}. 
Recently~\cite{kolchinsky2018semantic} a formalism for semantic information (SIT), was introduced using state spaces and probability distributions to assess mutual information between an agent and its environment, with persistence measured through a viability function. 
By scrambling information between systems of agents embedded in an environment, SIT quantifies how viability responds to altered information flow.

While promising, the formalism faces computational challenges, as it requires precise computation of high-dimensional state spaces and joint probability distributions. 
Thus, simplifying measures are needed for practical application, including effective definitions of viability, methods for scrambling information, and dimensional reduction techniques~\cite{pemartin2024shortcuts}. 
Recent studies applied SIT to models of foraging~\cite{Sowinski:2023vf} and synchronization~\cite{sowinski2024semantic}, capturing core biological behaviors like exploration, resource gathering and collective behavior, providing a practical and computationally feasible roadmap for applying SIT to real-world applications.

A key result from \cite{Sowinski:2023vf} was the identification of a viability plateau, where specific correlations between the agent and its environment had no measurable effect on survival, demonstrating that not all information is semantic. 
Beyond this threshold, increasing noise resulted in a decline in viability. This novel insight, which holds regardless of specific foraging strategies, highlights the potential for broader applications of SIT to other systems, including the context considered in this work. 

The SIT framework can be summarized in the following steps:
\begin{itemize} 
\item Decompose the system's coarse-grained degrees of freedom into Agent and Environment components. 
\item Identify and quantify the key biotic characteristics of the Agent that contribute to its viability. 
\item Quantify the correlations present in the system, specifically between the Agent and its Environment. 
\item Investigate and articulate the relationship between biotic characteristics and these correlations. \end{itemize}
The formalism also necessitates a causal analysis across all possible interventions. 
However, for all but the smallest systems, this approach is computationally infeasible.
Therefore, in this study, we omit the interventions step, focusing instead on clarifying the correlation structure between the Agent and Environment as a function of external conditions, such as variations in stellar luminosity.

\section{The exo-Daisy World Model}\label{sec: model}

In its simplest incarnation, Daisy World (DW) is a planet with a fraction of its area, $f$, habitable by two species of flora, black and white daisies, occupying, respectively, $f_B$ and $f_W$ fractions of the planets area. 
In what is to follow, the respective populations are indexed by $\alpha\in\{B,W\}$. 
The growth or decline of each flora is governed by a Malthusian law with a carrying capacity constraint, $f_W+f_B\le f$, thus,
\begin{align}
    \label{eq: DW EoM1&2}
    \frac{df_\alpha}{dt}=\beta(T_\alpha)(f-f_W-f_B)f_\alpha-\gamma_D f_\alpha.
\end{align}
Here $\gamma_D$ is a decay rate (the same for both species), and $\beta(T)$ the temperature dependent growth rate.
The latter peaks at some optimal temperature, $T_\text{opt}$, and has a characteristic width $\Delta T$.

The ground has an albedo $A_G$, while the flora have albedos $A_B<A_G$ and $A_W>A_G$.
Consequently, the planetary albedo is the weighted sum,
\begin{align}
    A=A_G+\sum_\alpha (A_\alpha-A_G)f_\alpha.
\end{align}
\emph{Rein control} is exerted by the flora through the positive and negative differences of each species' albedo with respect to the ground.
The planetary temperature is constrained by the solar flux impingent on the planetary surface.
For a stellar luminosity $L$, and planetary orbital radius $r$, this is given by the Stefan-Boltzmann law,
\begin{align}
    \label{eq: temperature constraint}
    T^4 = \frac{1-A}{16\pi\sigma r^2}L,
\end{align}
where $\sigma \simeq 5.6703\times 10^{-8}\text{ W}/\text{m}^2\text{K}^4$.
Due to their differing albedos, the flora are at different temperatures than the planetary average
\begin{align}
T_\alpha^4=T^4+q(A-A_\alpha),
\end{align}
which is governed by the coupling constant $q$. The magnitude of the differences in turn determines the growth rates of the respective species.

From a mathematical standpoint the DW model is a fully deterministic system governed by two equations of motion and a single constraint, Eq.~\eqref{eq: temperature constraint}. 
This constraint assumes that the thermal equilibration timescale is much shorter than the timescales associated with the growth and decline of the flora. 
As a toy model, however, it is not intended to directly represent the complexities of real planetary biospheres. 
For example, Earth experiences numerous stochastic events---such as volcanic eruptions and earthquakes---that are not captured in the deterministic framework of DW. 
While incorporating such real-world stochasticity presents significant challenges, one approach to introducing variability is by modeling the full coupling of star and planet on comparable timescales. 
This approach is motivated by M-dwarf planets, which present conditions where these timescales are more closely aligned. 
In this way, M-dwarf systems serve as a useful heuristic for breaking the constraint in Eq.~\eqref{eq: temperature constraint}.
By coupling these timescales, we elevate temperature and stellar luminosity to dynamical degrees of freedom in our exo-Daisy World (eDW) model. 
Fluctuations in stellar luminosity inject stochasticity into the system, providing a basis for the application of the SIT formalism to explore the informational dynamics of the model.

Considering M-dwarfs, we note that flaring is one mechanism for luminosity fluctuations  with over $20\%$ of M5 stars showing regular flaring events~\cite{martinez2020catalog}.
At even lower mass scales, $L$ dwarfs have variable brightness due to cloud formation on their surfaces. 
Using such variations as motivation for our toy model, we describe the luminosity of eDW's host star by an Ornstein-Uhlenbeck process,

\begin{align}
\label{eq: eDW EoM3}
    \frac{dL}{dt}=\frac{1}{\tau_s}\left(\langle L\rangle - L\right)+\sqrt{\frac{2}{\tau_s}}\delta\langle L\rangle \eta.
\end{align}
This ensures that the mean luminosity is $\langle L\rangle$ with fluctuations of order $\delta\langle L\rangle$ where $\delta\ll1$, and $\eta$ is a heuristic white noise.
The timescale of these stellar fluctuations is $\tau_s$.
We consider the exoplanet in orbit of this star to have an atmosphere of scale height $h\ll R_p$ (the radius of the planet), density $\rho$, and specific heat $c_V$.
A simple model for the planetary temperature dynamics assumes no work done on the atmosphere, resulting in 
\begin{align}\label{eq: eDW EoM4}
    \frac{dT}{dt}=\frac{1-A}{16\pi r^2 h  \rho c_V}L-\frac{\sigma}{h \rho c_V}T^4.
\end{align}
At equilibrium, this reproduces constraint Eq. \eqref{eq: temperature constraint}, and gives an equilibration timescale of $\tau_E=\rho h c_V/\sigma T_{eq}^3$.
Together Eqns~\eqref{eq: DW EoM1&2}, \eqref{eq: eDW EoM3} and \eqref{eq: eDW EoM4} form a coupled set of stochastic differential equations describing the dynamics of  eDW. The equations are closed with the provision of a growth rate.

\begin{figure*}
    \centering
    \includegraphics[width=0.75\textwidth]{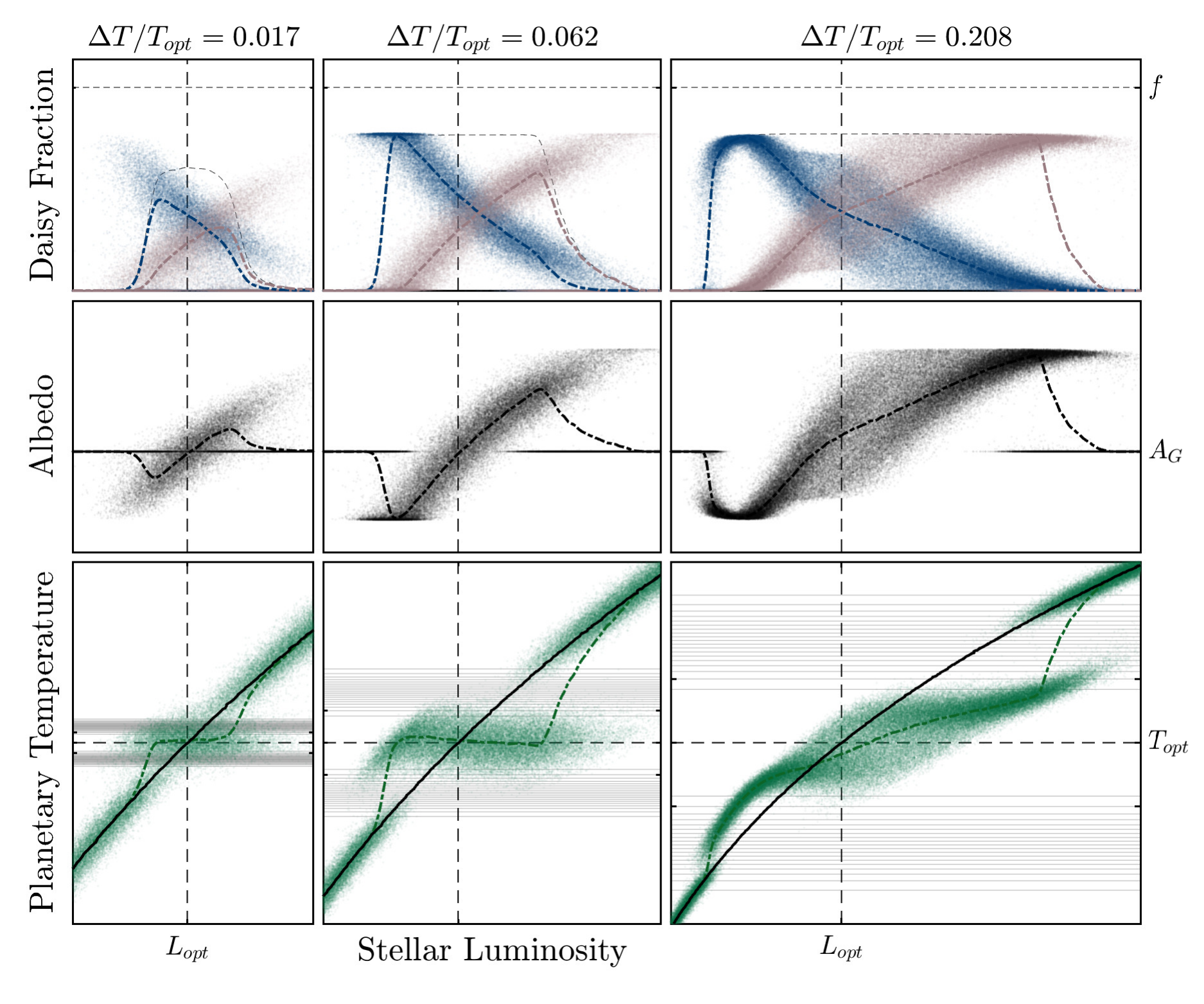}
    \caption{Qualitative behavior of the model; each column is an ensemble generated at a fixed daisy growth rate bandwidth as a fraction of the optimal temperature.
    (Top Row) Population sizes of the two species of Daisy at equilibrium over a range of stellar luminosities. 
    Scatter plots different realizations from an ensemble, dot-dash lines are mean values. (Middle Row) The Daisies alter the planetary albedo, causing (Bottom Row) the mean temperature of the planetary surface to get reined into the strip of temperatures that optimizes the daisy growth rate. 
    Horizontal dashed line is the optimum temperature, with light grey contours drawn at $5\%$ differences above and below. 
    Vertical dashed line in all plots is the stellar luminosity that gives the optimal temperature on a biome free planet. }
    \label{fig: FIG1}
\end{figure*}

In the original formulation of DW, the growth rate $\beta(T)$ was introduced with a quadratic dependence on temperature, as follows:
\begin{align}
    \beta(T) = \begin{cases}
        \gamma_G(1\!-4\!\frac{(T\!-\!T_\text{opt})^2}{\Delta T^2})& |T\!-\!T_\text{opt}|\!<\frac{\Delta T}{2}\\
        0 & \text{otherwise},
    \end{cases}
\end{align}
with $\Delta T$ the growth rate bandwidth (the range of bearable temperatures for the daisies), and $\gamma_G$ their optimal growth rate.
While this quadratic form is straightforward, in cases where differentiability is a concern, smooth functions like Gaussians have been employed as alternatives. 
When smoothness is not required, simpler forms such as triangular or rectangular window functions are often sufficient. For this study, we adopt a functional form that remains differentiable yet preserves a rectangular-like window shape:
\begin{align}
    \beta(T) = \gamma_G e^{-8\left(\frac{T-T_\text{opt}}{\Delta T}\right)^4}.
\end{align}
This formulation offers both the smoothness needed for differentiability and the sharp transitions characteristic of window functions. 

To better understand the timescales relevant to our stochastic eDW model, we provide some estimations that highlight why M-dwarf planets serve as a useful exemplar. 
These stars exhibit luminosities ranging from $10^{-4}$ to $10^{-1}L_\odot$ (where $L_\odot = 3.846 \times 10^{26}$ W is the solar luminosity), and their masses fall between $0.075$ and $0.6 M_\odot$\cite{henry2024character}. 
Exoplanets orbiting such stars typically have orbital radii of $0.05-0.25$ AU~\cite{tuomi2019frequency}, resulting in orbital periods as short as several days.
With planetary albedos assumed to be approximately $0.3$, the equilibrated average surface temperatures of these exoplanets could span a wide range, from $50$ K to $600$ K. 
Assuming atmospheric densities greater than Earth's ($1-10 \text{ kg}/\text{m}^3$), specific heat capacities between $10^3-10^4 \text{ J}/\text{kg K}$, and scale heights of $1-10$ km, the corresponding equilibration timescales range from a single day to $10^6$ years. 
For exoplanets with temperatures near $200$ K, this timescale is on the order of several days.
Given that the orbital period provides an annual cycle for exoplanets with axial tilt or orbital eccentricity, we argue that this cycle also defines the timescale for biosphere dynamics. 
In the eDW model, this implies that $\beta(T_\text{opt})^{-1} \sim \tau_E$, emphasizing the necessity of incorporating temperature dynamics into the model.
It is important to note that we are not suggesting that M-dwarf planets are expected to directly instantiate the eDW model; rather, we use these planets to illustrate the implications of timescale synchronization within the model.

Figure \ref{fig: FIG1} illustrates the behavior of eDW across a range of average stellar luminosities (x-axes) and growth rate bandwidths (columns).
Each point is taken from a single instance of the model; the description of the model dimensionalization, numerics, and ensemble statistics is fully detailed in Appendix \ref{app: numerics}.
The mean values of the daisy fractions, albedo, and planetary temperature (dashed lines in each of the panels) behave as in the original DW model.
The bandwidths of growth rate are depicted in the bottom panels, manifesting as horizontal contour lines centered around the optimal temperature, $T_{opt}$, with each contour corresponding to a successive 5\% decline in growth rate. 
As the system transitions beyond the tolerable thermal range, the daisy populations (top panels) collapse, and planetary temperature dynamics become governed primarily by the bare surface albedo (middle panels). 
Consequently, the planetary temperature converges according to the Stefan-Boltzmann law, as represented by the thick black curves in the bottom panels.

At the lower threshold of the viable temperature range, the black daisy populations (blue curves) predominate, leading to decreased planetary albedo, which in turn drives a thermoregulatory feedback loop that elevates the geosphere's temperature towards the optimal range. 
Conversely, at the upper thermal boundary, the white daisy populations (pink curves) become dominant, increasing planetary albedo, thereby inducing a cooling feedback mechanism that lowers the planetary temperature. 
This exemplifies the well-known \emph{rein control} mechanism, where the biosphere exerts regulatory influence on planetary conditions.

While the eDW behaves similarly to the classical DW model on average, its inherent stochasticity introduces far richer dynamics, particularly at the boundaries of the habitable temperature and luminosity range. 
The bottom panels of Fig.~\ref{fig: FIG1} illustrate the planetary temperature exhibiting fluctuations across all stellar luminosities.
For lower bandwidths (left and middle columns), these fluctuations remain confined within 5\% of the optimal temperature across the entire habitable luminosity spectrum.
However, for higher bandwidths (right column), significant edge effects emerge at both the cooler and warmer extremes of the viable temperature range. 
These edge effects are also evident in the daisy fraction and albedo (middle panels), stemming from the non-linearity of the Stefan-Boltzmann law, which becomes increasingly pronounced as the system explores a broader range of bearable temperatures.

\section{Informational Narrative}
\label{sec: info_n}

In advancing our analysis, we adopt a Semantic Information Theory (SIT) framework to characterize the system's complex behavioral dynamics. 
This approach necessitates partitioning the system into two distinct components: the first representing the agent and the second representing the environment. 
In its current formulation, the eDW model is governed by the evolution of four degrees of freedom, $(f_B, f_W, T, L)$, leading to 14 possible ways to partition these variables into non-empty agent and environment subsystems. 
Aligned with the Gaia hypothesis, the most natural partitioning is to designate the planetary biome as the agent, while the stellar and planetary parameters constitute the environment. 
Following this rationale, we define the agent's degrees of freedom as  as $\bs{a} = (f_B, f_W)$ and the environmental degrees of freedom as $\bs{e} = (T, L)$.

\begin{figure*}
    \centering
    \includegraphics[width=\textwidth]{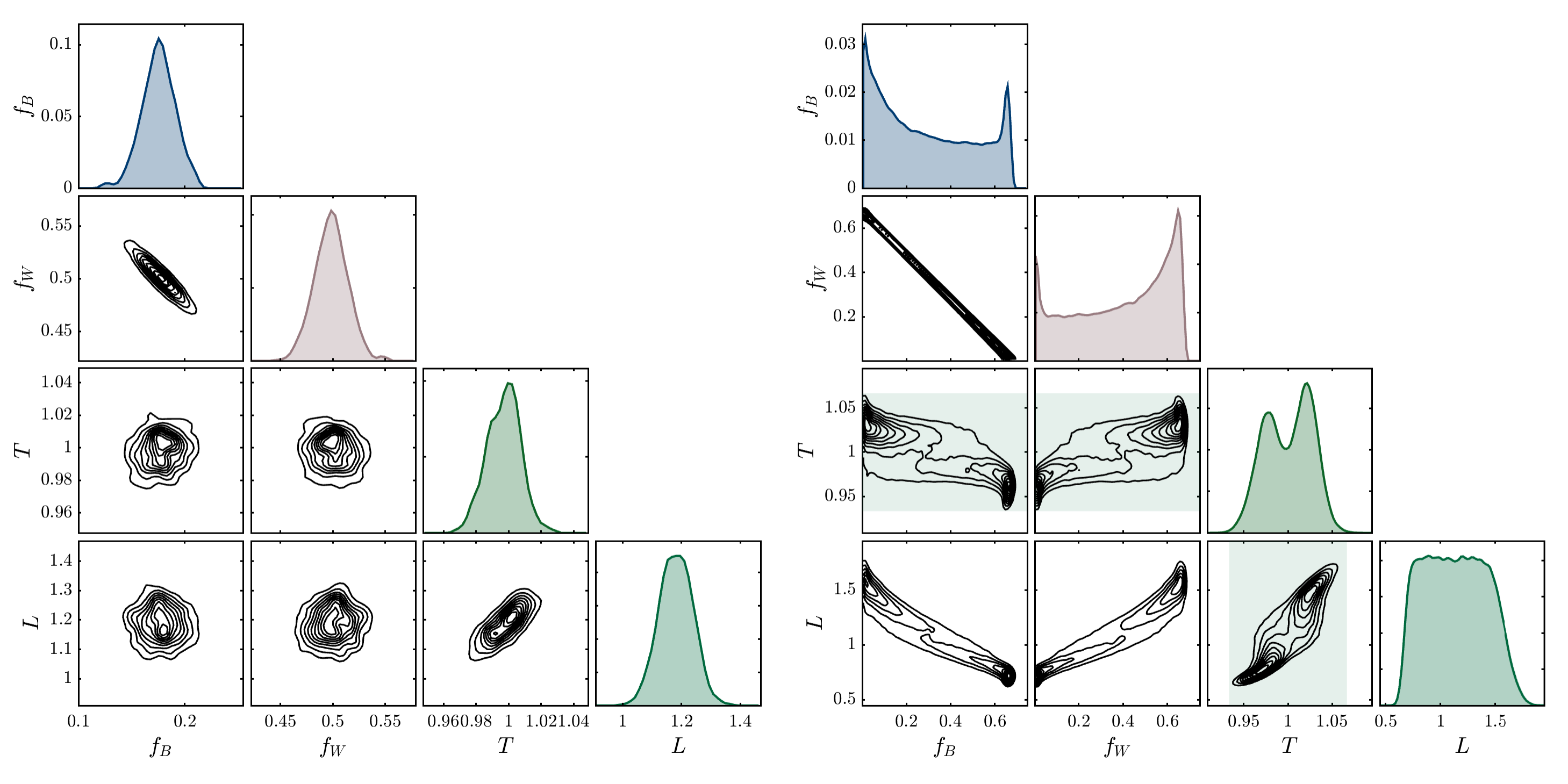}
    \caption{Distributions of DoF in the eDW model, and joint distributions between pairs of DoF. 
    (Left) Conditioned on an average stellar luminosity of $L = 1.1895L_\text{opt}$, and a bandwidth of $\Delta T = 0.0833 T_\text{opt}$. 
    It is clear that the environmental DoF are correlated with one another, as are the Agent DoF. 
    The pairwise correlations between Agent and Environment are not as clear. 
    (Right) Not conditioned on any average stellar luminosity, with a bandwidth of $\Delta T = 0.1667 T_\text{opt}$ (Visualized by light green strip in three of the subplots). 
    Over all luminosities, the pairwise correlations between Agent and Environment degrees of freedom are much clearer. }
    \label{fig: FIG2}
\end{figure*}

Next, we turn to the identification of viability.
Since this is intrinsic to the agent, it must depend explicitly on the agent degrees of freedom.
A natural choice for viability is the expected proportion of the habitable fraction of the planet occupied by the biome, given by,
\begin{align}\label{eq: viability} 
V = \mathds{E}^A\left[\frac{f_B + f_W}{f}\right], 
\end{align}
where $\mathds{E}^A$ represents the expectation value in the presence of the agent. 
The total biome area is normalized to the maximal habitable area for ease of interpretation, i.e, $0 \leq V \leq 1$, with $V = 1$ indicating that the biome occupies all of the habitable space on the planet.

We now focus on the correlations between the agent and environment, which can be extracted from the joint distribution 
$p_{AE} = p_{AE}(\bs{a},\bs{e} | \bs{\theta})$. Here, 
$\bs{\theta}$ epresents the full set of model parameters, as outlined in Appendix~\ref{app: simulation}, and includes both the average stellar luminosity, $\langle L\rangle$, and biome temperature bandwidth, $\Delta T$. It is important to note that the agent's viability, as defined in Eq. \eqref{eq: viability}, is implicitly dependent on both of these parameters.
Causal relationships between the agent and environment can also be determined by intervening on the system to obtain the agent-free distribution $p_0(\bs{e}) = p_{AE}(\bs{e}|f_B=f_W=0,\bs{\theta})$~\cite{pearl2009causality}.
From both $p_{AE}$ and $p_0$, the next step is to systematically enumerate and estimate the joint entropies for all possible subsets of degrees of freedom. 
Information measures are then derived from specific combinations of these entropies, which will be discussed in detail later, to capture the informational architecture that characterizes eDW's dynamics.
To avoid overwhelming the reader with extensive numerical details, the procedure for estimating both distributions is presented in Appendix~\ref{app: simulation}.

The left panel of Fig.~\ref{fig: FIG2} shows all the univariate and bivariate marginals of the estimated $p_{AE}$ from simulations with the average stellar luminosity fixed at $\langle L \rangle = 1.1895 L_\text{opt}$.
We chose this value in order to motivate the need for information measures when analyzing correlations. 
In the panel both intra-agent ($f_B$ vs. $f_W$) and intra-environment ($T$ vs. $L$) degrees of freedom exhibit clear covariance in their bivariate marginals.
However, the inter agent-environment degrees of freedom do not display as obvious a relationship; the bivariate distributions appear circular.
Two-point measures such as covariance have a hard time extracting inter-subsystems correlations in this setting.

On the right, we marginalize $p_{AE}$ over the history of the star's average stellar luminosity.
The luminosity evolution is modelled by a linear increase from $0.3 L_\text{opt}$ to $2.4 L_\text{opt}$; such increases in luminosity over main sequence evolution are typical of low-mass stars~\cite{salaris2005evolution}.
On the evolutionary timescale the agent-environment degrees of freedom display clear covariance.
Taken together with the left panel, the highly non-linear interactions between biome and environment generate three-point and higher correlations between the two on short timescales; two-point correlations become prominent only after integrating the system across evolutionary timescales.

Extracting correlations between more than two degrees of freedom on these short timescales necessitates the use of more advanced analytical tools, with information theory offering the required level of sophistication. 
The foundational tool within this framework is Shannon entropy. 
For a discrete random variable $X$ sampled from the distribution $p_X$, the entropy is defined as
\begin{align}
\label{eq: entropy} 
H(X)=-\sum_x p_X(x)\log_2 p_X(x) 
\end{align}
and is measured in bits.
Under the $\mathcal{Q}$-interpretation~\cite{gleiser2018configurational}, a bit corresponds to the average number of yes/no questions required to determine the exact value of the random variable, i.e., to identify that 
$X=x$. 
Under the $\mathcal{S}$-interpretation, entropy represents the uncertainty or \emph{surprise} associated with the outcome of $X$; an inevitable outcome carries no surprise. When
$X$ is unknown, there is a potential for surprise, and asking questions incrementally reduces that uncertainty until 
$X=x$ is known with certainty, at which point the initial surprise is replaced by a sequence of bits representing $x$. As surprise diminishes, information increases, a relationship that led Schr\"odinger and Brillouin to associate information with negative entropy\cite{schrodinger1944life,brillouin1953negentropy}.

Since $X$ is not restricted to being a scalar, entropy can be defined for collections of random variables. 
For a set of $n$ random variables, each of the $n!$ possible subsets has an associated entropy. These entropies can be combined in various ways to construct more complex information-theoretic measures, each offering deeper interpretative insights. 
One of the most fundamental of these measures is mutual information. 
Consider two random variables, $X$ and $Y$, and let their Cartesian product be denoted by  $XY$. 
The mutual information between the two variables is defined as
\begin{align}
    \label{eq: mutual information}
    I(X:Y) &= H(X)+H(Y)-H(XY)\nonumber\\
    &=\sum_{xy} p_{XY}(x,y)\log_2 \frac{p_{XY}(x,y)}{p_X(x)p_Y(y)}.
\end{align}

In the first expression in Eq.~\eqref{eq: mutual information}, we have demonstrated how the measure is constructed from the entropies discussed earlier. 
In the second, these entropies are reorganized in terms of the joint distribution and its marginals. 
The former provides an intuitive interpretation: $H(XY)$ is the number of bits needed to represent $x$ and $y$ together, while $H(X)+H(Y)$ is the number of bits needed to represent $x$ and $y$ independently. 
If fewer bits are required to represent $x$ and $y$ together than independently, this implies that knowledge of one variable reduces the uncertainty in the other---mutual information quantifies exactly how much information is shared between the two.
The latter expresses mutual information as a Kullback-Leibler divergence. 
Given that the Kullback-Leibler divergence is necessarily positive definite and measures the deviation from complete independence (as given by the product distribution in the denominator), it reinforces the interpretation of mutual information as a measure of shared information~\cite{kullback1951information}.
Since mutual information is sensitive to non-linear correlations between entire sets of degrees of freedom (DoF), it serves as an effective tool for capturing inter-agent-environment correlations, denoted $I(A:E)$. 
Additionally, one can also quantify intra-agent correlations  $I(a_1:a_2)$ and intra-environment correlations $I(e_1:e_2)$ .
All three measures are interpreted as quantifying the degree of dependency between the respective subsets of DoF.

A biome, by restructuring planetary energy flows to suit its own needs, exerts a causal effect on planetary-stellar correlations. 
The question arises: does the addition of a biome to a planet strengthen or weaken these pre-existing correlations? 
While the answer may vary depending on the specific case, the causal relationship between intra-environmental dependencies and the presence of an agent can be explored by computing the mutual information from the agent-free distribution, $p_0$. 
The quantity $I_0(e_1:e_2)$ quantifies the strength of the correlation between stellar luminosity and planetary temperature. 
In the limit where $\tau_E \ll \tau_s$, the planetary temperature responds almost instantaneously to stellar fluctuations, leading to a high level of dependency between the two, as knowledge of one almost completely determines the other.
Conversely, in the limit $\tau_E \gg \tau_s$, the slow response of the planet ensures that its temperature remains highly correlated with the average stellar luminosity, but not with short-term luminosity fluctuations. 

We focus on the dynamically interesting scenario where these two timescales, along with the agent's timescale, are comparable.
To quantify the agent's effect, we take the difference between $I_0(e_1:e_2)$ and $I(e_1:e_2)$:
\begin{align}
\label{eq: drop IE} 
 \Delta I_{\varnothing\rightarrow A} = I(e_1:e_2)-I_0(e_1:e_2),
\end{align}
which measures the production or destruction of intra-environmental correlations caused by the agent.

Unfortunately, we cannot conduct a similar causal analysis with the agent, as the question of what happens to a biome when its planet's star is removed has a simple and catastrophic answer. However, we propose a method to examine how the two daisy species cooperate with each other and the environment.
Letting $X$ and $Y$ be as before, and introducing a third random variable $Z$, the conditional mutual information is given by
\begin{align}
    I(X:Y|Z) &= I(X:YZ)-I(X:Z),
\end{align}
which represents the information shared between $X$ and $YZ$, minus the information shared between $X$ and $Z$ alone. 
It measures the information shared between $X$ and $Y$ that cannot be attributed to $Z$ alone. 
We utilize this to define intra-agent cooperation as influenced by the environment thus,
\begin{align}\label{eq: interaction information}
    C(a_1\!:\!a_2||E)&=  I(a_1\!:\!a_2)-I(a_1\!:\!a_2|E),\nonumber\\
    &=H(E)-H(a_1E)-H(a_2E)\nonumber\\ &\ \ \ \ 
    -H(A)+H(AE),
\end{align}
with the first line offering the seemingly straightforward interpretation of the information shared between the biome degrees of freedom (DoF) that can be solely attributed to the environment. 
Seemingly, because this definition can be reformulated into the manifestly symmetric form shown in the second line, which is known as interaction information~\cite{james2011anatomy, williams2010nonnegative}. 
Interaction information is known to take both positive and negative values. 
In the former case, it represents shared information across all random variables, interpreted as information redundantly distributed. 
Negative values of cooperation suggest a synergistic effect between the DoF, where the presence of the third variable enhances the correlation between the other two.

\section{Results}\label{sec: results}

\begin{figure}
    \centering
    \includegraphics[width=\columnwidth]{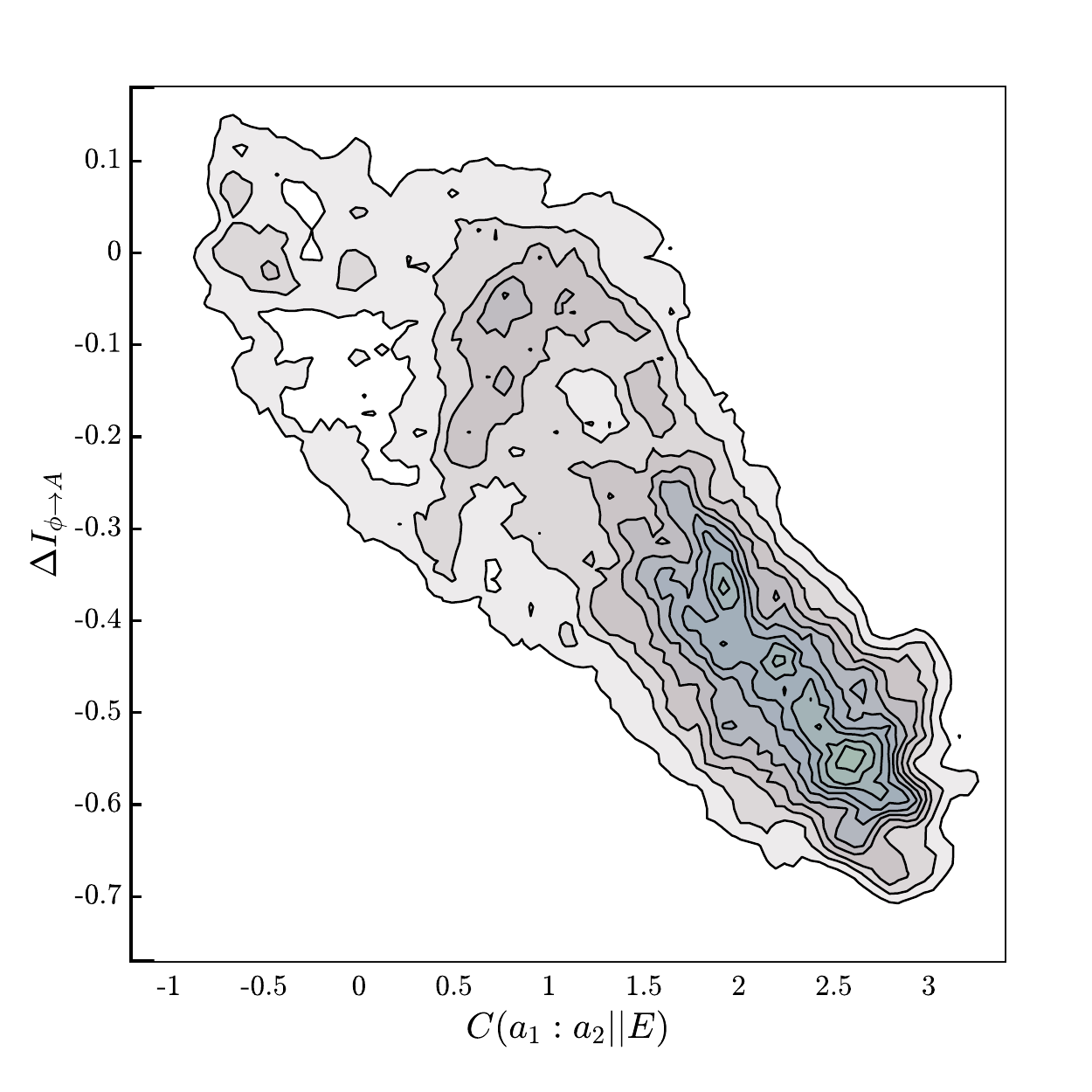}
    \caption{Kernel density of the distribution of cooperation between the biosphere DoF (Eq. \eqref{eq: drop IE}) and change in intra-environmental correlations (Eq. \eqref{eq: interaction information}) caused by the presence of the biosphere at high viability, $V>0.7$. 
    The drop in environmental correlation is strongly related to the robustness of the biosphere.}
    \label{fig: FIG3}
\end{figure}

The information-theoretic measures introduced in the previous section were not the only ones we analyzed; however, they were the ones that produced the most compelling results for constructing an informational narrative of eDW. 
Our first significant finding concerns the relationship between the change in intra-environmental correlation caused by the presence of the agent, Eq. \eqref{eq: drop IE}, and the cooperation between individual agent DoF and all environmental DoF, Eq. \eqref{eq: interaction information}. 
Figure \ref{fig: FIG3} presents the kernel density computed across all simulations with high viability, specifically those with $V > 0.7$. 
The values on the y-axis are predominantly negative, indicating that the introduction of the biome tends to decorrelate the environmental degrees of freedom from each other. 
In contrast, the x-axis values are mainly positive, suggesting that the cooperation between the daisy fractions creates redundant correlations with the environment. 
Taken together, the figure reveals how correlations are managed by the biome, i.e., the informational architecture of eDW. 
Maintaining high viability is closely tied to the biome's ability to convert environmental correlations into redundant correlations with itself. 
The more the planetary temperature becomes decoupled from the stellar luminosity, the more constrained the agent's degrees of freedom become. 
This finding elegantly explains the edge effects observed in Figure \ref{fig: FIG1}, where planetary temperature fluctuations become much more constrained at the cooler and warmer boundaries of the bearable temperature range.

\begin{figure}
    \centering
    \includegraphics[width=\columnwidth]{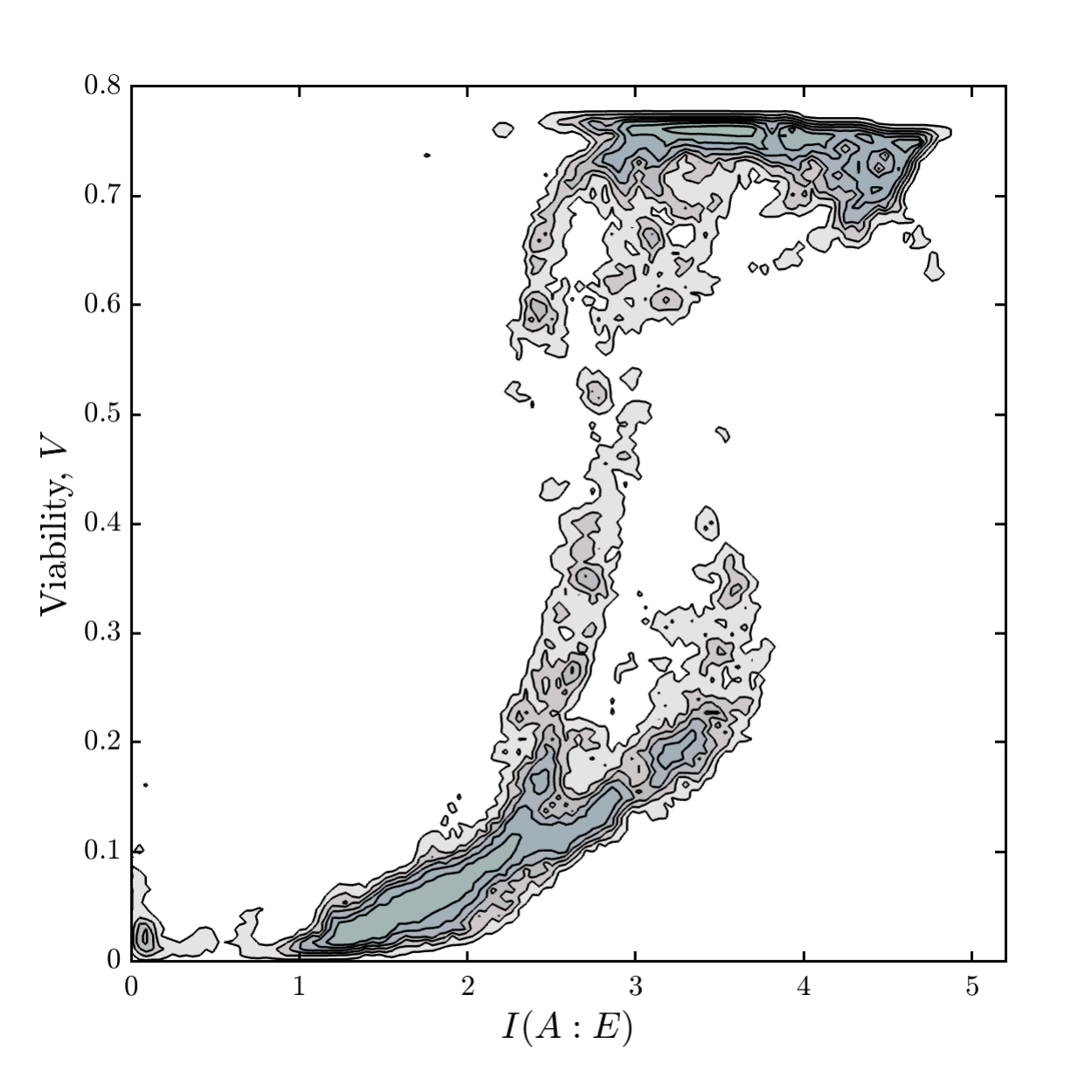}
    \caption{ Kernel density of the distribution of viability and total inter agent-environment correlations. Most of the data points are in the two clusters observed, so in order to accentuate structure the contours are logarithmically spaced.}
    \label{fig: FIG4}
\end{figure}

Our second finding, highlighted in the introduction, is the canonical SIT viability curve shown in Figure \ref{fig: FIG4}. 
The horizontal axis represents the mutual information, Eq. \eqref{eq: mutual information}, between the full set of agent DoF and environmental DoF, while the vertical axis displays the viability, Eq. \eqref{eq: viability}. 
The kernel density reveals distinct clustering in the upper-right and lower-left regions of the parameter space. 
These clusters align with the viability curve predicted by semantic information theory \cite{kolchinsky2018semantic}. 
The cluster in the upper-right forms a viability plateau, indicating a set of agent phase space trajectories that achieve maximal viability, unperturbed by additional correlations with the environment.
As we move towards the left, the sharp decline in viability reflects an information threshold, analogous to the minimal amount of agent-environment correlation required to sustain viability, as described in SIT. 

We use the term analogous intentionally here, as there appear to be two sharp increases in viability, occurring at around 2.5 and 3.5 bits. 
It is important to note that the precise numerical values are not significant, as they depend on the binning used to estimate the joint distribution. In this case, the maximum mutual information is approximately at $\log_2 25\approx 4.64$ bits, so the observed thresholds occur at roughly $\sim 0.5$ and $\sim 0.75$ of the maximum.
It is possible that these thresholds represent a form of hysteresis, where the higher threshold corresponds to the minimum correlation required to establish a viable biome, and the lower threshold corresponds to the minimum correlation needed to sustain that biome. 
While we do not assert the presence of a strict semantic threshold due to the absence of a full counterfactual analysis, we recognize both the thresholds and the viability plateau as important features of eDW's informational architecture.

\section{Discussion} \label{sec: discuss}

Daisy World is a simplified model illustrating how a biosphere exerts homeostatic control over its host planet in response to changes in an external astrophysical driver, such as increasing stellar luminosity. 
As such, it has played a critical role in debates surrounding Gaia Theory and the extent to which biospheres can co-evolve with their planets to provide regulatory feedback that maintains habitable conditions. 
Initially developed to study Earth's history, Gaia Theory's implications have gained renewed relevance in astrobiology. 
ExoGaia models propose that long-term habitability necessitates sustained inhabitation, meaning that Gaian feedback mechanisms must be established for biospheres to persist on geological timescales. 
This longevity is crucial for astronomers, as it increases the likelihood of detecting biosignatures across interstellar distances. 
Thus, Daisy World models, as fundamental tools for exploring Gaian feedback, remain relevant decades after their introduction.

In this work, we introduce a novel approach to Daisy World models using Semantic Information Theory, a framework that allows us to explore what we refer to as \emph{informational narratives} in living systems. 
SIT focuses on correlations between an agent and its environment that contribute to the agent's viability.
In this paper, we introduced a simplified version of the SIT framework to develop a Daisy World model capable of incorporating information-theoretic measures, such as mutual information. 
Our exo-Daisy World model assumes that stellar luminosity fluctuates stochastically at the few percent level over timescales comparable to the growth rates of daisies. 
While this assumption situates the model within parameter regimes reminiscent of M-dwarf habitable zone planets, this association is not essential to the conclusions of our study.

The stochastic eDW model allowed us to investigate the information, or correlational structure, within and between the biosphere and planetary/stellar degrees of freedom (DoFs). 
The biosphere's DoFs are represented by the two daisy populations, while the latter DoFs include temperature and stellar luminosity. 
In simulations of planets lacking biospheres, the joint DoF distributions revealed strong correlations between temperature and luminosity, consistent with the Stefan-Boltzmann law. 
However, with the inclusion of a biosphere, these correlations between planetary DoFs begin to break down as stellar luminosity increases. 
As daisy species proliferate, the biosphere effectively \emph{steals} correlations from the environmental DoFs by exerting rein controls over them. 
These changes reflect the flow of information between the biosphere and planetary system. 
We defined interaction information as a measure of cooperation between daisy species. 
This allowed us to identify a semantic threshold in the biosphere-environment dynamics, which marks the total mutual information load necessary for the biosphere to maintain viability over timescales matching the stellar luminosity increase.

While homeostatic feedback in Daisy World is typically described in terms of physical quantities (e.g., radiation fluxes, albedos, and coverage fractions), our work offers a complementary perspective. 
Alongside the physical narrative, our study demonstrates how Daisy World evolution can be recast as an informational narrative. 
Information flows between the environment and the biosphere, with the biosphere "using" that information to establish correlations that enhance its viability by regulating the environment. 
Crucially, not all of this information is meaningful for maintaining control. 
By analyzing the structure of information-theoretic quantities (e.g., joint distributions, mutual information, and interaction information), we can identify where, how, and how much of this information contributes to the biosphere's viability and resilience.

Given that Daisy World is fundamentally a toy model, we do not claim that our analysis reveals new dynamic features. 
Rather, our work highlights the potential of applying information-theoretic approaches like SIT to models of co-evolution between biospheres and their planets. 
More realistic models would need to account for the complex networks of interactions between the living and non-living planetary subsystems. 
These include the atmosphere, hydrosphere, cryosphere, lithosphere, and most critically, the biosphere, all of which exert intricate webs of feedback on the planetary system. 
Because the biosphere, as a living system, processes information in ways non-living systems do not, information-theoretic approaches can capture aspects of coupled system dynamics that might otherwise remain invisible to purely physical or chemical models.

Furthermore, similar to the application of network-theoretic measures \cite{10.1093/mnrasl/slad156}, if informational narratives prove essential in describing biosphere-planetary control (i.e., Gaian or semi-Gaian control), they could lead to the identification of new classes of agnostic biosignatures. 
As a result, the next step in our research program will involve applying SIT and other information-theoretic approaches to more complex models of coupled planetary systems.

\section{Acknowledgement}  
The authors thank the Center for Integrated Research Computing (CIRC) at the University of Rochester for providing computational resources and technical support. 
This project was partly made possible through the support of Grant 62417 from the John Templeton Foundation.
The opinions expressed in this publication are thoseof the authors and do not necessarily reflect the views of the John Templeton Foundation.

\vspace{10px}
\appendix
\section{Simulation Methodology}\label{app: simulation}
\subsection*{Dimensionalization}
Exo-Daisy World is a dynamical system of four degrees of freedom, $ \{f_B,f_W,T,L\}$, whose evolution is described by
\begin{widetext}
\begin{align}
    \frac{df_B}{dt}&=\beta(T_W)(f-f_B-f_W)f_B-\gamma_D f_B\\
    \frac{df_W}{dt}&=\beta(T_B)(f-f_B-f_W)f_W-\gamma_D f_W\\
    \frac{dT}{dt}&=\frac{1}{16\pi r^2 h  \rho c_V}L(1-A_G-(A_B-A_G)f_B-(A_W-A_G)f_W) -\frac{\sigma}{h \rho c_V}T^4\\
    \frac{dL}{dt}&=\frac{1}{\tau_s}(\langle L\rangle -L)+\sqrt{\frac{2}{\tau_s}}\delta\langle L\rangle \eta\\
    &\hspace{30pt}\text{where}\hspace{20pt} \beta(T) = \gamma_G e^{-\left(\frac{T-T_\text{opt}}{\Delta T}\right)^4}\\
    &\hspace{35pt}\text{and}\hspace{20pt}  T_{\alpha}^4=T^4+q(A_B-A_G)f_B+q(A_W-A_G)f_W - q(A_{\alpha}-A_G)
\end{align}
\end{widetext}
The $16$ parameters of varying dimensions, $\{f,A_G,A_B,A_W,q,\gamma_G,\gamma_D, \tau_*,r,h,\rho,c_V,T_\text{opt},\Delta T,$ $\delta,\langle L\rangle \}$, fully specify the model.
Dimensionalizing reduces this parameter space, greatly simplifies the equations of motion, and makes the system amenable to numerical simulation, which this appendix covers.

To this effect we scale all temperatures to $T_\text{opt}$, then define the optimal luminosity $L_\text{opt}=16\pi r^2\sigma T_\text{opt}^4/(1-A_G)$, and use it to scale luminosity.
This lets us parameterize the average stellar luminosities as $\langle L \rangle = (1+\lambda)L_\text{opt}$, where $\lambda\in(-1,\infty)$.
The environmental timescale is $\tau_E = c_V\rho h/\sigma T_{\text{opt}}^3$.
Time is scaled to the biosphere timescale $\gamma_G^{-1}$.
Lastly we introduce the dimensionless growth window
\begin{align}
    w(x) = e^{-\alpha x^4},
\end{align}
where $\alpha = 8 T_\text{opt}^4/\Delta T^4$, the scaled temperature deviation $Q=q/T_\text{opt}^4$, and the shorthand $\delta A_\alpha = A_\alpha -A_G$.

The dimensionalized equations of motion are now
\begin{widetext}
\begin{align}
    \label{eq: eDW1}
    \frac{df_B}{dt}&=w(T_W-1)(f-f_B-f_W)f_B-\frac{\gamma_D}{\gamma_G} f_B\\
    \label{eq: eDW2}
    \frac{df_W}{dt}&=w(T_B-1)(f-f_B-f_W)f_W-\frac{\gamma_D}{\gamma_G} f_W\\
    \label{eq: eDW3}
    \frac{dT}{dt}&=\frac{1}{\gamma_G\tau_E}\left(1-\frac{\delta A_Bf_B+\delta A_Wf_W}{1-A_G}\right)L -\frac{1}{\gamma_G\tau_E}T^4\\
    \label{eq: eDW4}
    \frac{dL}{dt}&=\frac{1}{\gamma_G\tau_s}(1+\lambda -L)+\sqrt{\frac{2}{\gamma_G\tau_s}}\delta(1+\lambda) \eta\\
    &\hspace{30pt}\text{where}\hspace{20pt}  T_{\alpha}^4=T^4+Q\delta A_Bf_B+Q\delta A_Wf_W - Q\delta A_{\alpha}
\end{align}
\end{widetext}
We collect the remaining $11$ dimensionless parameters and write them as the parameter vector 
\begin{align}
    \bs{\theta}^T\!\!\!=\!
(f,A_G,\delta\! A_B,\delta\! A_W,Q,\frac{\gamma_D}{\gamma_G}, \gamma_G\tau_E,\gamma_G\tau_s,\alpha,\delta, \lambda )
\end{align}
and the dimensionless DoF as the vector 
\begin{align}
    \bs{x}^T = (f_1,f_2,T,L).
\end{align}
This prescription makes it apparent that eDW is a vector process described by the stochastic differential equation
\begin{align}\label{eq: SDE}
    d\bs{x} = \bs{a}(\bs{x}|\bs{\theta}) dt + \bs{b}(\bs{x}|\bs{\theta}) dW
\end{align}
where $W$ is a one-dimensional Wiener process.

\subsection*{Numerical Integration}\label{app: numerics}
Standard numerical integration methods such as Euler or Runge-Kutta must be modified non-trivially for stochastic differential equations (SDE)\cite{sarkka2014lecture}.
These modifications can make a stochastic RK2 method significantly more complicated, and tedious, to implement than a much higher non-stochastic RK method\cite{rossler2009second}. 
Fortunately for us, Eq.~\ref{eq: SDE} does not have the general form of a multivariate SDE --- the Wiener process is scalar --- allowing us to use a strong RK1 method \cite{roberts2012modify}

\begin{algorithm}[H]
\caption{Stochastic Runge Kutta\\ of Strong Order 1}\label{alg:cap}
\begin{algorithmic}
\Require $\bs{X}(t),\bs{\theta},\Delta t$
\Ensure $\bs{X}(t+\Delta t)$
\State $\bs{X}\gets \bs{X}(t)$
\State Draw $Z\sim \mathcal N(0,1)$
\State Draw $S\sim \mathcal U\{\pm 1\}$
\State $\bs{k}_1 \gets \bs{a}(\bs{X},\bs{\theta})\Delta t + (Z-S)\bs{b}(\bs{X},\bs{\theta})\sqrt{\Delta t}$
\State $\bs{k}_2 \gets \bs{a}(\bs{X}+\bs{k}_1,\bs{\theta})\Delta t + (Z+S)\bs{b}(\bs{X}+\bs{k}_1,\bs{\theta})\sqrt{\Delta t}$
\State $\bs{X}(t+\Delta t)\gets \bs{X}+\frac{1}{2}(\bs{k}_1+\bs{k}_2)$
\end{algorithmic}
\end{algorithm}

\subsection*{Statistics}
The dimensionless parameters chosen for all the simulations are in the upper part of table $\ref{table: model parameters}$.
The parameters that differ for each simulation, the average stellar luminosity, $(1+\lambda)L_\text{opt}$, and growth rate bandwidth, $\Delta T$, are in the lower half of the table.
The stellar luminosities span an order of magnitude from $0.3 L_\text{opt}$ to $2.4 L_\text{opt}$, while the bandwidths span two orders of magnitude from $0.002 T_\text{opt}$ to $0.267 T_\text{opt}$.
The former range is sampled at $400$ evenly spaced luminosities and the latter at $128$ evenly spaced bandwidths; resulting in $51,200$ simulations, each at a fixed average stellar luminosity and growth rate bandwidth.
An additional $400$ simulations are run, one at each fixed stellar luminosity with the agent free model.

Each simulation samples the joint distribution, $p_{AE}$ or $p_0$, by running $500$ independent instances of eDW with the following initial conditions:
For the former distributions, $f_B$ is chosen uniformly at random in the range $[0,f]$ for each instance, and then $f_W=f-f_B$ is set.
For the latter distributions both $f_B=f_W=0$ are set.
The luminosity is initialized to the average stellar luminosity, and the temperature is initialized to the corresponding equilibrium temperature, Eq. \ref{eq: temperature constraint}.
A dimensionless timestep of $dt=0.1$ is used to integrate each instance $2000$ times, using the strong stochastic Runge-Kutta method of order $1$ described earlier, then recording the values of the four/two DoF.
This procedure ensures that any transients have decayed away and that the individual recordings making up the ensembles are not correlated.

\begin{table}[t!]
\caption{Simulation Dimensionless Parameters}
\centering 
\begin{tabular}{lcr}    \toprule
\emph{Parameter} & \emph{Symbol} & \emph{Value} \\\midrule
 Habitable Land Fraction   & $f$  & $0.88$  \\ 
 Ground Albedo      & $A_G$  & $0.3$  \\
 Black Daisy Albedo Difference & $\delta\! A_B$  & $\mhyphen 0.2$\\
 White Daisy Albedo Difference & $\delta\! A_W$ & $0.3$ \\
Temp. Shift Heat Constant & $Q$ & $0.1$ \\
Decay to Growth Rate Ratio &$\gamma_D/\gamma_G$ &0.2\\
Environmental Timescale & $\gamma_G \tau_E$ & 5.0\\
Stellar Timescale &$\gamma_G\tau_s$&3.0\\
Stellar Fluctuation Scale &$\delta$ & 0.05\\
\midrule
Average Stellar Luminosity      & $\lambda$ & $\mhyphen 0.7-1.4$ \\
 Growth Rate Bandwidth  & $\Delta T/T_\text{opt}$ & $0.002\!-\!0.267$ \\
 \bottomrule
 \hline
\end{tabular}\label{table: model parameters}
\end{table}

Each ensemble is then used to estimate the corresponding joint distribution.
Since the DoF take on real values, we discritize their ranges using a modified square root rule: given $N$ samples, we subdivide the corresponding range into $1+\lceil\sqrt{N}\rceil$ equally sized bins.
For the agent-free ensembles this means the temperature and luminosity ranges, 
$[\min(T),\max(T)]$ and $[\min(L),\max{L}]$ respectively, are divided into $1+\lceil \sqrt{500}\rceil =25$ bins.
For $p_{AE}$, the number of instances, $N$,  with $f_B>0$ and $f_W>0$ is used to subdivide the four analogous ranges (which now include the pair $[\min(f_\alpha),\max(f_\alpha)]$) into $1+\lceil \sqrt{N}\rceil$ bins.
The empirical distributions are then constructed by counting the number of instances in each bin, and normalizing. 
These are our estimates of the joint distributions, which are used to compute the expectation value needed to estimate the viabiity, eq. \ref{eq: viability}, as well as our corpus of information measures, namely $I(A:E),\Delta I_{\phi\rightarrow A},$ and $C(a_1:a_2||E)$. 
Every marginal of the joint distributions is found, and the corresponding naive estimator of the marginal entropy computed. 
These are then used to construct estimates of all the information measures used in the analysis.

All simulations were run in parallel on the Blue Hive cluster at the Center for Integrated Computational Research (CIRC) at the University of Rochester. 
Since the instances within the ensembles were run in series, total computation time per simulation ranged from approximately $10$ minutes to $8$ hours.
Analysis was performed on a $2020$ Macbook Pro with Apple's M1 processor, using MATLAB~\cite{MATLAB} and Mathematica~\cite{Mathematica}.
\end{document}